# Determination of thermal and optical parameters of melanins by photopyroelectric spectroscopy

J. E. de Albuquerque[a)]
*Departamento de Física, Universidade Federal de Viçosa, Viçosa, 36571-000, MG, Brazil*

C. Giacomantonio, A. G. White, P. Meredith
*Soft Condensed Matter Physics Group, School of Physical Sciences, University of Queensland, Brisbane, QLD 4072, Australia*

Photopyroelectric spectroscopy (PPE) was used to study the thermal and optical properties of electropolymerized melanins. The photopyroelectric intensity signal and its phase were independently measured as a function of wavelength, as well as a function of chopping frequency for a given wavelength in the saturation part of the PPE spectrum. Equations for both the intensity and the phase of the PPE signal were used to fit the experimental results. From the fittings we obtained for the first time, with great accuracy, the thermal diffusivity coefficient, the thermal conductivity and the specific heat of the samples, as well as a value for the condensed phase optical gap, which we found to be 1.70 eV. © *2005 American Institute of Physics*. [DOI:

The melanins are a class of bio-macromolecule found throughout nature.[1] In humans, they act as pigments and photoprotectants in our hair, skin and eyes. They are also found in the brain stem and inner ear, where their roles are less well defined. Eumelanins (also known as dopa-melanins), the predominant form of the macromolecule in humans, are broad band ultra-violet and visible light absorbers. Additionally, they are the only known solid state bio-macromolecule semiconductor.[2] This combination of properties has led to the novel proposition that they may be useful from a technological perspective as functional "electronic soft-solids".[1] Despite significant scientific effort over the past 30 years, large knowledge gaps concerning the basic physics and chemistry of melanins still exist. These materials are particularly intractable from the analytical perspective, since they are chemically and photochemically very stable, and virtually insoluble in most common solvents. Hence, we do not fully understand key properties such as electronic structure vs. chemical composition, or why (and how) melanins conduct electricity in the condensed solid state. Significant knowledge gaps also exist regarding melanin biofunctionality. It is well accepted that these molecules serve as our primary photoprotectants. However, the mechanisms by which melanin aggregates dissipate potentially harmful solar radiation are not well understood. It has been proposed that efficient coupling between photo-excited electrons and phonon modes allows non-radiative dissipation of absorbed photon energy – i.e. melanins can effectively turn biologically harmful photons into harmless heat.[3]

Photothermal spectroscopic (PTS) techniques have been extensively and successfully applied to solid state materials for obtaining their thermal and optical parameters.[4–6] In these techniques a pulsed light beam is absorbed in a solid sample and the converted heat (non-radiative conversion) diffuses into the bulk structure; the sample expansion, or the temperature gradient, is then detected by an appropriate sensor system. The detected signal depends on the optical and thermal properties of the sample: the optical absorption coefficient $\beta(\lambda)$ ($\lambda$ being the light wavelength), the non-radiative conversion efficiency $\eta(\lambda)$, the thermal conductivity $k$, and the thermal diffusivity coefficient $\alpha$. The signal also depends on experimental control parameters such as the chopping frequency $f$ of the incident light beam. Among the PTS techniques, photoacoustic spectroscopy (PAS),[7] which is the most traditional one, and the more recent photopyroelectric spectroscopy (PPES) have been used for studying thermal and optical properties of polymeric films.[8–11] Among the various physical parameters which can be measured, the thermal diffusivity is particularly important because this measurement allows one to obtain the thermal conductivity and specific heat. Furthermore, the importance of this measurement resides in the fact that, as with the optical absorption coefficient, the thermal diffusivity is unique for each material. When the thermal conductivity is known, information can be obtained regarding the heat transfer process by phonons and by carriers (electrons or holes). In this current study, we used samples of electropolymerized (EP) melanins on indium tin oxide (ITO) glass and compressed powder as self supporting pellets. Equations for both the intensity and the phase of the PPE signal, taking into account the thermal and the optical characteristics of the pyroelectric detector, have been used to fit the experimental results.

The detected signal $V(\omega,t)$, $\omega=2\pi f$, is proportional to the pyroelectric coefficient $p$ of the detector and to the temperature distribution along the detector thickness:[8–13]

$$V(\omega,t) = \left[ \frac{p}{K\varepsilon_0} \int_{L_p} T_P(\omega,x)\,dx \right] e^{i\omega t} \quad (1)$$

where $L_p$ is the sample thickness, $T_p(\omega, x)$ the temperature of the sample, $K$ the dielectric constant of the material, $i = (-1)^{1/2}$, and $\varepsilon_o$ the vacuum dielectric permittivity. The heat propagation across the whole chamber is governed by heat diffusion equations of each medium coupled via boundary conditions at the interfaces ($T_a = T_b$ and $k_a dT_a/dx = k_b dT_b/dx$, $a$ and $b$ representing consecutive media), as established by Mandelis and Zver.[12] The signal $V(\omega,t)$ obtained by integrating the diffusion equations is normalized by the ratio $V(\omega,t)/V_R$, where $V_R$ is the signal measured directly over the detector painted with a very thin layer of a black ink. In this latter case, the detector is considered thermally thick and optically opaque, and the normalized voltage signal as given in Refs. 8–13.

a) Author to whom correspondence should be addressed,
currently visiting The University of Queensland,
electronic mail: jeduardo@ufv.br.

Assuming the case where the sample is in an optically opaque condition, that is, in the saturated region of the spectra, then the normalized voltage and phase signal can be expressed as:[9–11]

$$V_n(\omega) = 2\eta_s (b_{gs} + b_{ps}) \div [(b_{gs}+1)(b_{ps}+1)e^{\sigma_s L_s} - (b_{gs}-1)(b_{ps}-1)e^{-\sigma_s L_s}] \quad (2)$$

$$F_n(\omega) = -\arctan[\gamma \cdot \tan(a_s L_s)] \quad (3)$$

where:

$$\gamma = \frac{(b_{gs}b_{ps}+1)\cosh(a_s L_s) + (b_{gs}+b_{ps})\sinh(a_s L_s)}{(b_{gs}+b_{ps})\cosh(a_s L_s) + (b_{gs}b_{ps}+1)\sinh(a_s L_s)} \quad (4)$$

and where $\sigma_n = (1+i)a_n$ with $a_n = (\pi f/\alpha_n)^{1/2}$ ($n = g, s, p, b$, that is, $g$=gas, $s$=sample, $p$=pyroelectric, $b$=backing), and $b_{nm} = k_n a_n / k_m a_m$.

The photothermal spectrometer used in our experiments is schematically shown in Fig. 1. It comprises an optical part (light source, monochromator, and chopper), the custom-made pyroelectric or photoacoustic chamber, and the measuring system. When the monochromator is used, either an optical cable can be connected to the photothermal chamber, substituting the mirror, or the chamber can be put directly into exit slit of the monochromator (Jobin Yvon HRS 2). The measuring system is composed of a Stanford Research System SR 530 lock-in amplifier, locked at the chopper frequency, and connected to a pc that stores the data and controls the experiment. The mechanical slotted wheel chopper (SR 540) modulates the incident light. The nonradiative conversion efficiency $\eta(\lambda)$ is very near to unity, because luminescence effects in melanins have very low efficiency (less than $10^{-3}$).[14] The PPE cell was set up with a silica window above the sample for the melanin pellets, and for the EP films the window was the ITO glass substrate itself. The wavelength range was between 350 and 1100 nm. We used long pass filters to avoid overtones of the monochromator grating (this is extremely important for $l > 700$ nm). For frequency scanning experiments, we used a green He:Ne laser centered at 543.5 nm (Melles Griot 05-LGR-193) as a power source, and for wavelength scanning we used a 150W ozone free Xenon arc lamp (Thermo Oriel).

Thin films of dihydroxyphenylalinine (DOPA)-melanin were synthesised by oxidative electropolymerisation of DL-DOPA (Sigma-Aldrich). Initial solutions were 30mM of DL-DOPA in sodium tetraborate buffer (Sigma-Aldrich, 0.1 M, pH 9). Electrical current was passed through the solutions by dropping up to 20 V across a copper cathode and ITO anode using a DC power supply. To accelerate the formation of melanin, the solution was initially oxidised by mechanical stirring for 10 to 15 minutes at a voltage that generated 10-20 mA/cm$^2$ of current. The solution was then left in atmospheric conditions at a current density of 0.5 mA/cm$^2$ for 1–8 days, depending on the desired thickness of the film. During this time, the solution turned black and a soft black melanin film formed on the anode. Once the desired thickness was achieved, films were dried slowly in a sealed container with various saturated salt solutions to control the relative humidity. The humidity was stepped down from 94% to about 50% over a period of 3 to 5 days. Slow drying minimised cracking of the films. In the course of drying, the films decreased in thickness from about 100-500 μm to 1–2 μm for the thinnest film. We obtained samples with various thicknesses, ranging from ~1 to 65 μm. Synthetic powders of DOPA-melanin were extracted from the black, electropolymerised DL-DOPA solutions by acidification to pH 2.0 with 6 M hydrochloric acid. The precipitated melanin was separated from the solution by centrifugation at 3500 rpm for 10 minutes and then dried in air. Powders were pressed at 400 MPa into pellets 192-500 μm thick, with the thinnest pellets most suitable for PPE measurements.

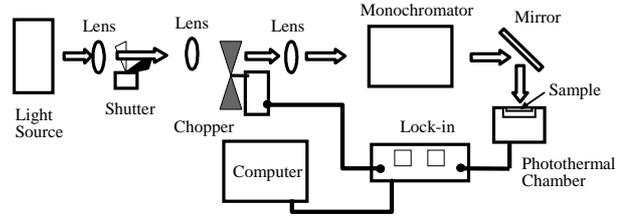

FIG. 1. Experimental setup of the acquisition data system.

Figures 2 and 3 show, respectively, the normalized phase $F_n(f)$ and voltage amplitude $V_n(f)$ as a function of the chopping frequency, for the melanin pellet. At the wavelength of the green power source ($\lambda = 543.5$ nm), the sample is highly absorbing as shown in Fig. 4. The data were recorded in the saturated region of the PPE spectra, for the case where the detector is thermally thick, that is, above 10.7 Hz. The experimental points for the normalized phase obey a linear dependence on the square root of the frequency for frequencies below 120 Hz (see Fig. 2). This means that the fractional term $\gamma$ (Eq. (4)) in Eq. (3), is approximately unity in this frequency range. In fact, $\gamma$ differs from unity by less than 1 %, when we consider appropriate values for thermal conductivity and diffusivity coefficients. This permits us to approximate Eq. (3) to the simple relation $F_n \cong a_s L_s$. As such, the thermal diffusivity $\alpha_S$ is directly obtained from the slope of the fitting curve $F_n$ vs. $f^{1/2}$ (the continuous line of Fig. 2), using the relation $a_S = (\pi f/\alpha_S)^{1/2}$, and its value is shown in Table I.

Table I. Physical parameters of the melanin pellet.

| | |
|---|---|
| $L_s$ (μm) | 192 ± 5 |
| $\rho_s$ (g/cm$^3$) | 1.43 ± 0.01 |
| $\alpha_s$ (m$^2$/s) | $(2.96 \pm 0.05) \times 10^{-7}$ |
| $k_s$ (W/m K) | (0.106 ± 0.002) |
| $c_s$ (J/kg K) | (250 ± 7) |

The $\alpha_s$ values of Table I were then used for the $V_n(f)$ fitting utilizing Eq. (2), and the thermal conductivity $k_s$ became the single adjusted parameter of the results shown in Fig. (3). The specific heat of the sample $c_s$ is directly derived from the relation $k = \rho c \alpha$, where $\rho$ is the mass density, valid for a stationary state. The values of $k_s$ and $c_s$ found via this analysis are also shown in Table I.

Figures 4 and 5 show PPE spectra at 20 Hz chopping frequency, normalized voltage and phase, respectively, of the 65 μm EP melanin film. The PPE $V_n$ signal follows approximately the optical transmission spectrum of the material, i.e., transmission-like, but saturates for wavelengths $\lambda$ below 730 nm. The $F_n$ spectrum of Fig. 5 shows a behavior consistent with the fact that the sample is optically opaque below 730 nm. In this case, the phase lag is greater in the region with higher opacity, and the normalized phase spectrum follows an absorption behavior as depicted in Fig. 5. Hence, from these spectra, we can conclude that the optical gap of the EP melanin starts at around 730 nm, i.e., 1.70 eV.

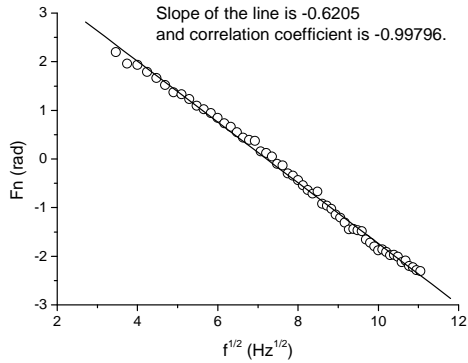

FIG. 2. Experimental points (dots) and line of best fit for the PPE normalized phase.

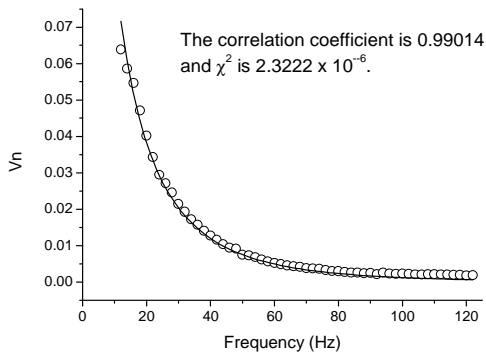

FIG. 3. Experimental points (dots) and line of best fit for the PPE normalized voltage.

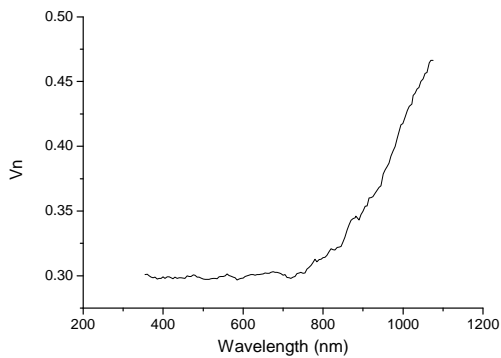

FIG. 4. PPE normalized voltage $V_n$ spectrum of the 65 µm EP melanin film.

In conclusion, in this paper we have presented the application of a particular photothermal technique, photopyroelectric spectroscopy, in thermal and optical studies of melanins. Despite the complexity of the PPE equations, establishing the optically opaque condition makes the theoretical approach a realistic tool to fit the experimental curves. The values given in Table I for a melanin pellet show a thermal diffusivity and a thermal conductivity near to that of insulating polymers, but a significantly lower specific heat. More importantly, our data indicates that these melanin samples possess a solid state optical gap of 1.70 eV. This value corresponds to the minimum energy required to cause a transition between the highest occupied and lowest unoccupied molecular orbitals (HOMO and LUMO respectively) of the system. As such, it corresponds to the HOMO-LUMO gap and is consistent with our first principles density functional theory calculations of the gap of indolequinone and hydroxyindole oligomers.[15] We believe our measurements to be the first direct observation of the optical gap of melanin, and its determination will undoubtedly assist in on-going effort to understand the condensed phase physics and chemistry of these important bio-materials.

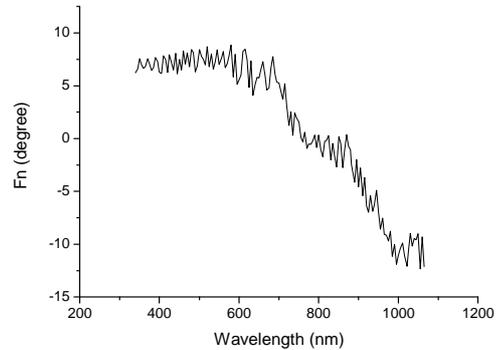

FIG. 5. PPE normalized phase $F_n$ spectrum of the 65 µm EP melanin film.

The first author acknowledges UFV for sabbatical leave. This work was partially funded by the Australian Research Council (DP0345309) and by the MURI Center for Photonic Quantum Information Systems, ARO/ARDA Program No. DAAD19-03-1-0199.